\documentclass[conference]{IEEEtran}
\IEEEoverridecommandlockouts
\usepackage{cite}
\usepackage{amsmath,amssymb,amsfonts}
\usepackage{algorithmic}
\usepackage{graphicx}
\usepackage{textcomp}
\usepackage{xcolor}
\def\BibTeX{{\rm B\kern-.05em{\sc i\kern-.025em b}\kern-.08em
    T\kern-.1667em\lower.7ex\hbox{E}\kern-.125emX}}

\usepackage{hyperref} % For hyperlinks
\usepackage[bottom]{footmisc}
\usepackage{dirtytalk}

\usepackage{subcaption}
\usepackage{soul}

\begin{document}

\title{Optimizing Checkpoint-Restart Mechanisms for HPC with DMTCP in Containers at NERSC}

\author{
\IEEEauthorblockN{William Arndt\IEEEauthorrefmark{0}, Johannes P. Blaschke\IEEEauthorrefmark{0}, Lisa Gerhardt\IEEEauthorrefmark{0}, Madan Timalsina\IEEEauthorrefmark{0}$^{*}$, and Nicholas Tyler\IEEEauthorrefmark{0}}
\thanks{$^*$Corresponding author: mtimalisna@lbl.gov}
\IEEEauthorblockA{Lawrence Berkeley National Laboratory, Berkeley, CA 94720 USA\\
}
}

\maketitle
\thispagestyle{plain}
\pagestyle{plain}

\begin{abstract}
This paper presents an in-depth examination of checkpoint-restart mechanisms in High-Performance Computing (HPC). It focuses on the use of Distributed MultiThreaded CheckPointing (DMTCP) in various computational settings, including both within and outside of containers. The study is grounded in real-world applications running on NERSC Perlmutter, a state-of-the-art supercomputing system. We discuss the advantages of checkpoint-restart (C/R) in managing complex and lengthy computations in HPC, highlighting its efficiency and reliability in such environments. The role of DMTCP in enhancing these workflows, especially in multi-threaded and distributed applications, is thoroughly explored. Additionally, the paper delves into the use of HPC containers, such as Shifter and Podman-HPC, which aid in the management of computational tasks, ensuring uniform performance across different environments. The methods, results, and potential future directions of this research, including its application in various scientific domains, are also covered, showcasing the critical advancements made in computational methodologies through this study.
\end{abstract}


\section{Introduction}
The integration of Distributed MultiThreaded CheckPointing (DMTCP)~\cite{Arya_2014} across a spectrum of HPC environments marks a significant evolution in the computational sciences. DMTCP's adaptability, functioning both within and external to containerized frameworks, showcases its versatility across different HPC applications. This pioneering exploration into DMTCP's role within varied computational settings lays the groundwork for a deeper understanding of the checkpoint-restart (C/R) process, especially within the cutting-edge NERSC Perlmutter supercomputing ecosystem. NERSC Perlmutter stands at the forefront of scientific discovery, offering exceptional computational capabilities to a wide array of research initiatives.

This work aims to underline the practical utilities of checkpoint-restart mechanisms for managing complex computational tasks and highlight their enhanced efficiency and reliability in HPC workflows. The ability to periodically save and recover the state of running processes presents a crucial innovation in computational methodology, enabling the continuation of computations after interruptions and thus safeguarding valuable computational time and resources. By leveraging DMTCP both within and outside containerized environments, this study illustrates the significant benefits of C/R mechanisms. These include improved job scheduling flexibility, minimized computation restart times, and enhanced overall system resilience. Through this comprehensive examination, we aim to contribute to the ongoing advancement of computational practices. This ensures that the scientific community can continue to tackle the most challenging and complex problems with increased efficiency and reliability.

%%%%%%%%%%%%%%%%%%%%%%%%%%%%%
\section{Checkpoint-Restart}
\label{sec:cr}
The checkpoint-restart (C/R) mechanism plays a crucial role in HPC, aimed at enhancing the reliability and efficiency of large-scale computations. This method involves periodically saving the state of a running process or a set of processes, referred to as a \verb|checkpoints|. These checkpoints comprehensively capture the entire state of the process at a specific moment, including memory, current executing instructions, Input/Output (I/O) status, and other related data of the running process into a file. The primary objective of this process is to provide a recovery point. This allows the system to restart the computation from the checkpoints in case of a failure or batch scheduler interruption (in the case of HPC), rather than starting a new one from the beginning or scratch. This approach is particularly valuable in HPC environments due to complex and time-consuming computations. It facilitates job preemption by opportunistically utilizing spare CPU cycles and aids in the efficient scheduling of multicore and single-threaded jobs. It can significantly reduce application startup times and facilitate batch scheduler optimizations, including preemption, thereby avoiding the high costs of restarting~\cite{Paul_H_Hargrove_2006, Arya_2014, Egwutuoha2013, jia2022software}. 

The C/R strategy extends beyond HPC to cloud computing, providing fault tolerance for distributed applications. It is implemented across various operating systems and programming languages, establishing itself as a foundational element for resilient and adaptable computing in diverse and demanding computational landscapes~\cite{azeem2023performance, Rodriguez-Pascual2019, nersc-doc}. 
 
For HPC centers like NERSC, C/R allows strategic job management by preempting less urgent tasks in favor of critical and time-sensitive ones. This approach can optimize computational resources and improve node utilization. It enhances the cluster's overall efficiency and throughput by strategically backfilling smaller jobs around larger reservations. It enables system maintenance with minimal disruptions by allowing tasks to be paused and resumed as well as significantly improves the system reliability, especially against unforeseen power outages or hardware failures. In addition, it offers benefits to HPC users by extending job runtimes, smoother workflow management, and more effective debugging capabilities by restarting from specific saved states.

However, there are challenges to C/R technologies in HPC, primarily the complexity it introduces for users. The goal of user transparent C/R demands considerable effort and coding infrastructure to enable users to seamlessly pause and resume their computational tasks without complications. Furthermore, adding MPI (Message Passing Interface) support creates another layer of complexity, as integrating diverse MPI implementations and networks, leads to compatibility issues often referred to as the MxN problem. DMTCP (Distributed MultiThreaded CheckPointing) addresses these challenges, offering a way to simplify the user experience and enhance the robustness of HPC operations by managing the C/R process across different platforms~\cite{zvi2021optimized, 9139610, nersc-doc}.

\section{Distributed MultiThreaded CheckPointing (DMTCP)}
Distributed MultiThreaded CheckPointing (DMTCP) is an advanced tool designed to facilitate automatic and efficient checkpointing and restarting of multi-threaded and distributed applications within HPC environments. Notably, DMTCP operates in a transparent manner, capturing the entire state of a program—including all threads, memory, and open files—without the necessity for any modifications to the program's source code~\cite{Arya_2014, dmtcp-doc}. This feature proves particularly beneficial for complex HPC applications where altering the code to incorporate checkpointing functionalities is impractical or infeasible.

Moreover, DMTCP exhibits a high degree of versatility and user-friendliness, aligning it with a wide range of applications within the HPC domain. The system's adaptability is evident, seamlessly integrating with a variety of distributed computing environments. This includes compatibility with the Message Passing Interface (MPI), various programming languages like C, C++, Python, and Fortran, as well as shell scripts. Additionally, it efficiently interacts with different resource managers, including Slurm, enhancing its utility across diverse computing packages. The tool's proficiency in managing checkpointing in multi-threaded environments is critical for maintaining performance and ensuring reliability within HPC systems. DMTCP's robust checkpointing capability is vital for the continuation of long-duration computations, offering resilience against system failures and providing the flexibility to pause, migrate, or resume computations across different machines or environments as required~\cite{ansel2009dmtcp}.

At NERSC, DMTCP's reliable C/R mechanisms facilitate pausing, migrating, and resuming long-running computations, extending computational job walltimes through a preemptable queue. This strategy addresses the needs of urgent and real-time computing and maximizes the utilization of computational resources, facilitating a more adaptable deployment of the supercomputing infrastructure. The reliable C/R functionality also enables workflows to maintain their state beyond the constraints of a batch-scheduler “job”. This investigation elucidates the manner in which DMTCP assists NERSC users in the development of workflows that more effectively leverage the center's HPC resources~\cite{bard2022lbnl,9651304,doi_johannes}.

\subsection{How does DMTCP work?}

\begin{figure}[h]
	\centering
    \includegraphics[width=0.45\textwidth]{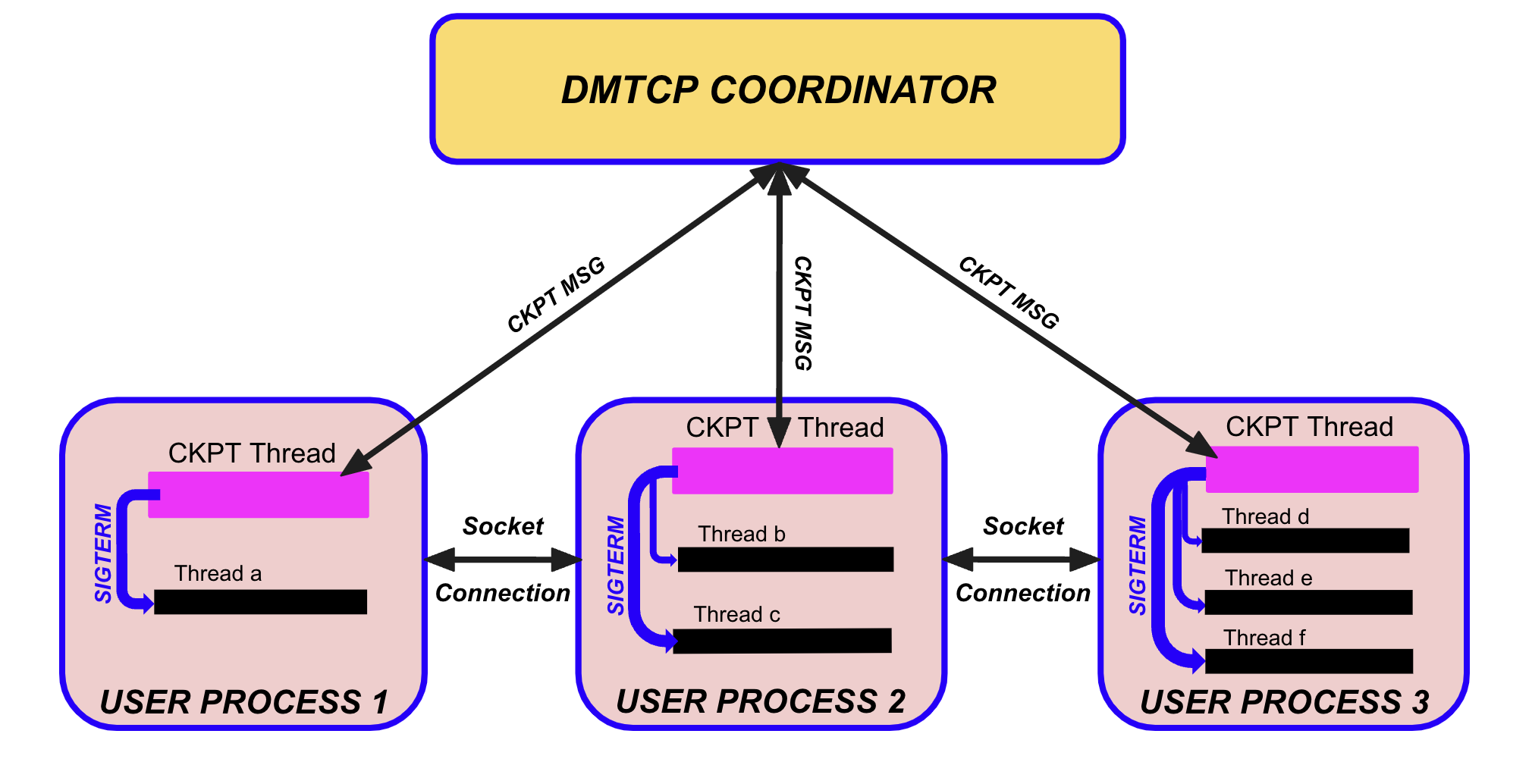}
     \caption[]{Diagram illustrating the Distributed MultiThreaded CheckPointing (DMTCP) system with a central coordinator managing checkpoint messages (CKPT MSG) with three user processes. Each process contains a checkpoint thread (CKPT Thread) and user threads (Thread a/b/c/d/e/f), interconnected via socket connections. Signals (SIGTERM) are also shown, indicating the communication between threads and the checkpointing mechanism. Upon receiving a CKPT MSG from the central coordinator, the checkpoint threads trigger a signal to user threads, and a checkpointing action is initiated, which involves saving the current state of the processes.}
    \label{DMTCP_artch_p}
\end{figure}

DMTCP employs a coordinator-based mechanism pivotal for managing checkpointing within HPC environments. The central coordinator (top part of Fig.~\ref{DMTCP_artch_p}) initiates the checkpointing process, establishing a one-to-one mapping between computational tasks requiring checkpointing and their dedicated management. With the support for multiple coordinators, the architecture enables independent, parallel checkpointing processes. Specialized threads execute checkpointing within user processes through socket-based communication with the coordinator~\cite{cao2016systemlevel}.

The functionality of DMTCP is further extended through a plugin architecture, which facilitates event hooks and function wrappers for process virtualization. This aspect is crucial for applications that need consistent system resources after a restart. The system uses virtual process IDs to manage resources such as file descriptors and network sessions transparently. DMTCP enhances fault tolerance and the system's ability to recover from coordinator failures without losing the runtime context by wrapping system and library calls, thereby supporting additional instrumentation and specialized setup during testing~\cite{7776536, chouhan2021architecture}.

The design emphasizes fault tolerance by redundantly storing checkpoint images and capturing the state of runtime libraries and environment variables. This approach guarantees that applications can resume operations post-restart with the same runtime context, including library APIs and modifiable environment settings. DMTCP plugins facilitate this seamless transition and recovery process.

\section{National Energy Research Scientific Computing Center (NERSC)’s HPC Containers}
HPC containers provide an efficient, scalable, and portable solution for executing complex applications. Containers encapsulate software, libraries and dependencies, ensuring uniform behaviour across different environments. They effectively tackle software compatibility challenges and simplify deployment, facilitating easier sharing and replication of work among researchers and developers~\cite{10.1007/s11227-022-04848-y}.

NERSC has introduced {\tt shifter} and {\tt podman-hpc}, specialized container technologies designed to address the rigorous demands of HPC applications, which include security, filesystem performance, robust communication, and optimized libraries. These solutions emphasize NERSC's commitment to enhancing application portability and reproducibility in varied computing environments. By encapsulating software dependencies, {\tt shifter} and {\tt podman-hpc} ensure consistent operation across different infrastructures, streamlining the process of sharing and replicating scientific work. This aligns with NERSC's objective of supporting cutting-edge scientific research by providing adaptable and user-centric computing resources, thereby facilitating a collaborative scientific community. Both {\tt shifter} and {\tt podman-hpc} are integrated with NERSC's HPC infrastructure, offering solutions that mitigate the limitations of traditional container technologies in HPC settings. These tools demonstrate significant advantages in terms of performance, security, and flexibility, making HPC containerization more accessible and effective for researchers and developers.

\subsection{Shifter}

{\tt shifter}, is a highly specialized containerization solution designed specifically for HPC environments. It offers an effective bridge between the Docker ecosystem and HPC systems, allowing for the seamless execution of Docker containers within HPC infrastructures. By leveraging {\tt shifter}, users can bring their Docker or other containerized applications directly to HPC resources without sacrificing performance or compatibility. This integration is crucial for exploiting the full capabilities of HPC systems while maintaining the ease of use and portability that Docker provides~\cite{nersc-doc, opensource}.

The development of {\tt shifter} addresses a critical need within the HPC community for a tool that can manage user-defined images with minimal overhead, thus ensuring that the sophisticated hardware of HPC systems is utilized efficiently. {\tt shifter} not only supports Docker images but also allows for their adaptation and optimization for HPC environments, incorporating features like high-speed networks and parallel file systems. This adaptation ensures that the containerized applications perform optimally on HPC systems, bridging the gap between traditional container usage and the specialized requirements of HPC workloads. Importantly, {\tt shifter} enhances the reproducibility and portability of scientific computing and research, key factors in accelerating scientific discovery.

The operational process of the {\tt shifter} system within the NERSC HPC infrastructure follows a user-focused and efficient workflow. It begins with users selecting or creating a Docker container image containing the necessary software and dependencies, which is uploaded to DockerHub. Upon accessing a Shifter-enabled HPC resource at NERSC, users issue a command to pull this Docker image, initiating its retrieval and preparation for use within the {\tt shifter} environment. Users then submit a batch job specifying the Docker image they wish to use, which executes within the container’s environment using commands such as aprun or srun. {\tt shifter} enhances flexibility by allowing the definition of volume mappings, enabling easy linking of external directories at NERSC to directories within the container, facilitating output file management. Additionally, containers configured with a specific entry-point can be run directly by users, bypassing the need for a detailed batch script and simplifying the focus on the computational task. The process concludes with users receiving the standard output and exit status, preserving the conventional experience of running batch jobs in an HPC context and ensuring that users can leverage the advantages of containerized applications within the performance and security frameworks typical of HPC environments~\cite{Jacobsen2015Contain}.

\subsection{Podman-hpc}
%%%%%
Podman (Pod manager), developed by Red Hat, is an open-source containerization tool that provides a secure alternative to Docker. It is fully compatible with Open Container Initiative (OCI) images and registries. Podman stands out for its daemonless architecture, which diverges from traditional container engines by eliminating the need for root privileges, thereby enhancing security. This structure allows users to run containers in rootless mode, significantly reducing the risk of security breaches that could compromise the host system~\cite{RedHat, oci}.

At NERSC, the {\tt podman-hpc} add-on specifically addresses the performance and usability challenges inherent in HPC applications. It enhances the performance and scalability of containerized HPC workflows, making it particularly effective for large-scale computations. This contrasts with other container solutions, such as {\tt shifter}, which, while providing similar capabilities, does not allow for dynamic modification of container contents at runtime. {\tt podman-hpc} empowers users with the ability to adjust container contents during runtime and to build images directly on the NERSC’s Perlmutter supercomputing system.

Users should generate a ``Containerfile" or ``Dockerfile" to begin working with {\tt podman-hpc}. A Containerfile is a more general form of a Dockerfile—they follow the same syntax and usually can be used interchangeably. Users can build and tag the image in the same directory via a command like: \verb|podman-hpc build -t elvis:test .|. To make an image suitable for use in a job, one must execute the command \verb|podman-hpc migrate elvis:test|, converting the image into a squashfile format compatible with {\tt podman-hpc}. These images can then be directly accessed and used in a job, or even pushed to an external registry like Docker Hub and pulled later as needed~\cite{nersc-doc, 10030014}.

Moreover, {\tt podman-hpc} facilitates the direct execution of containers via the \verb|podman-hpc| command, eliminating the need for an intermediary repository. This feature streamlines the deployment process and aligns with the unique security and performance requirements of HPC environments at NERSC. Users can pull public images using \verb|podman-hpc| with no additional configuration. Images pulled from a registry are automatically converted into a suitable squashfile format for {\tt podman-hpc} and can be accessed and used in a job directly. For pulling images from a private registry, users must first log in to their registry through \verb|podman-hpc|. Importantly, {\tt podman-hpc} supports both interactive and batch job execution without requiring special resources, provided the images have been built or pulled using \verb|podman-hpc|~\cite{nersc-doc}.

Furthermore, NERSC’s integration of { \tt podman-hpc} marks a major advancement in HPC tools, including Cray MPI and NVIDIA CUDA, which are crucial for parallel computing and advanced data processing tasks. The unique capabilities of {\tt podman-hpc}, such as runtime modification of containers and local image building, set it apart from other container technologies. Additionally, its ability to directly execute containers without an intermediary repository makes it a preferred choice. These features make {\tt podman-hpc} especially suitable for researchers and developers in security-conscious and performance-oriented HPC environments~\cite{10030014,presentation1}.

\subsection{Performance Benchmarking of HPC Containers at NERSC}
%%%%%%
\begin{figure}[h]
	\centering
    \includegraphics[width=0.45\textwidth]{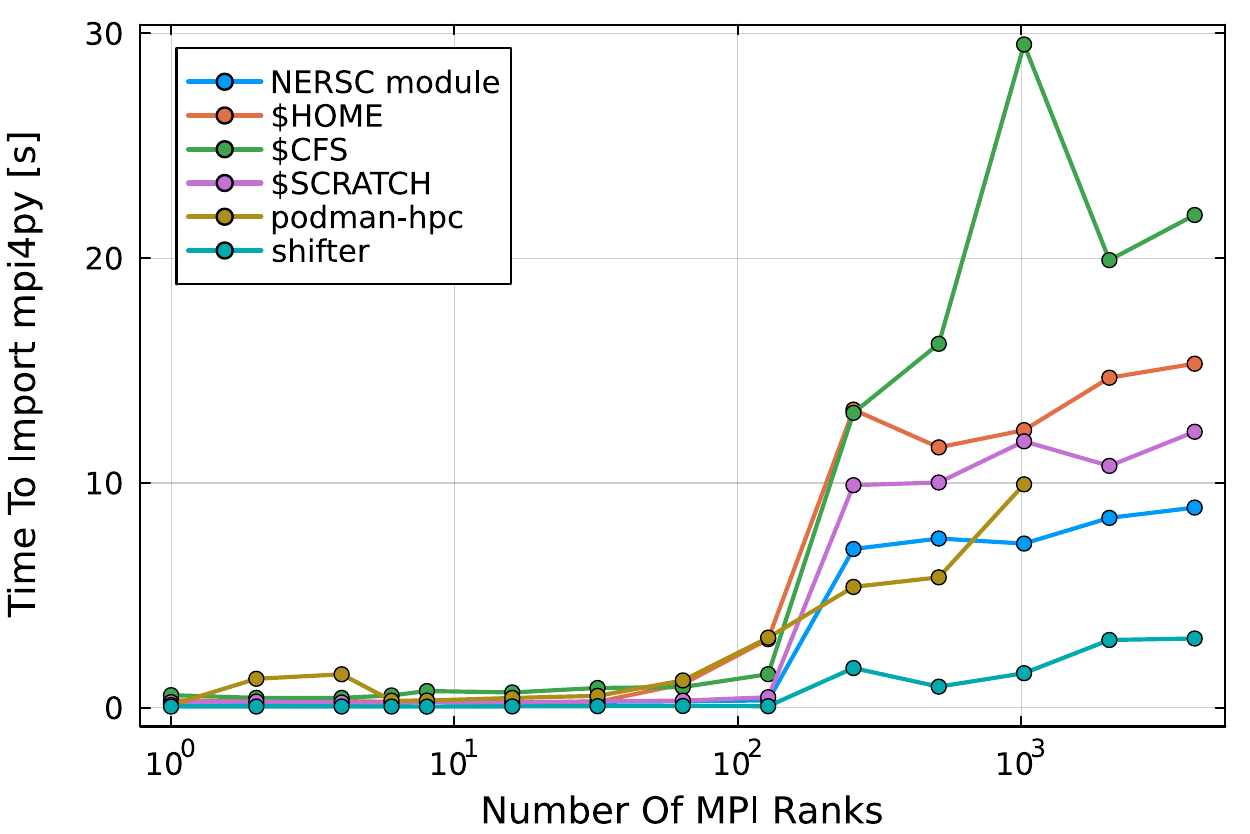}
     \caption[]{
        Mean execution time of {\tt from mpi4py import MPI}  as a function of number of MPI ranks, and location of Python environment (based on benchmark from~\cite{9651304}). Lines represent mean over multiple runs and ranks. This benchmark is collected on a Perlmutter CPU node, with up to 128 ranks per node. Correspondingly, we see that import times rapidly at around 128 ranks. Colored lines represent different file systems that the Python environment is located on. The ``NERSC module'' is installed to {\tt /global/common/software}, which is optimized to allow for highly parallel loading and linking of shared libraries. {\tt shifter} and {\tt podman-hpc} correspond to the two container runtime environments available on NERSC's Perlmutter system. {\tt podman-hpc}'s performance at scale is comparable with the highly-optimized file systems (HOME, SCRATCH, and {\tt /global/common/software}), whereas {\tt shifter} out-performs all others.
     }
    \label{podman_bench}
\end{figure}

Fig.~\ref{podman_bench} reveals the impact of which file system (or container runtime) a software package is installed to on startup time. This benchmark was collected on NERSC Perlmutter CPU nodes. The application benchmarked is a simple {\tt mpi4py} program in an Anaconda environment. Lines show the mean time to execute the statement {\tt from mpi4py import MPI} (the full code of the benchmark can be found in~\cite{9651304}). We see that an increase in the number of MPI ranks results in a longer time for Python to load (and dynamically link) all {\tt mpi4py} dependencies (the sudden rise in load time at 128 ranks corresponds to going from single node to multiple nodes).

Notably, Containers' role in reducing the wall clock time for the import and initialization is critical in parallel computing. Fig~\ref{podman_bench} shows that the configuration using Container, outperforms applications installed on shared file systems. We observe that {\tt shifter} performs better than {\tt podman-hpc} at scale, this is likely a result of {\tt podman-hpc} not having had the benefit of years of performance optimization (being a relatively new container runtime compared to {\tt shifter}).

The ability of the container runtime to cache images effectively reduces the time taken to load libraries dynamically, a common bottleneck in large-scale data processing, and it appears to be equally effective in smaller-scale environments.

Furthermore, this strategic adoption not only optimizes startup performance but introduces a level of operational resilience crucial for managing complex computational tasks. Incorporating DMTCP-based C/R mechanisms within containers can significantly improve the adaptability and efficiency of scientific workflows. It ensures that scientific investigations proceed smoothly, with minimal downtime and enhanced data processing capabilities. Containers like {\tt shifter} and {\tt podman-hpc} prove to be instrumental in this context, offering scalable solutions that support both extensive and nuanced research activities. The use of such containerized C/R frameworks underscores a significant shift towards more dynamic and reliable scientific computing environments, where the ability to quickly adapt and efficiently process data can substantially accelerate the pace of scientific discovery and innovation.
%%%%%%%%%%%%%%%%%%%%%%%

\section{Methods}
Our simulation experiments on NERSC Perlmutter, utilizing DMTCP, explored checkpoint-restart (C/R) processes both inside and outside containers. Our analysis encompassed both single-threaded and multi-threaded Geant4~\cite{AGOSTINELLI2003250} simulations across different versions and packages, providing a comprehensive understanding of their performance in different computational environments.

\subsection{On NERSC Perlmutter}
On NERSC's Perlmutter system, the C/R jobs submission and 
resubmission process is significantly streamlined through 
the CR Module (\verb|nersc_cr|), which utilizes DMTCP. This 
module includes a pivotal function, \verb|start_coordinator|, 
which activates the checkpointing mechanism via the 
\verb|dmtcp_coordinator| command. It sets the necessary 
environment variables for the coordinator's communication 
and manages the \verb|dmtcp_command.<jobid>| file, thereby 
facilitating efficient job coordination. As users initiate 
the job submission by submitting their script, the batch 
system dynamically allocates nodes by seeking backfill 
opportunities within the job's specified time constraints. 
When a job nears its time limit, it receives a USR1 signal, 
prompting the \verb|func_trap| function to execute the 
checkpoint command and requeue the job, thereby updating 
its remaining walltime. This mechanism ensures that the job 
cycle of execution, checkpointing, and requeuing persists 
until completion or the achievement of the desired duration, 
optimizing compute resource usage and enhancing job management 
efficiency.

To further augment this efficiency, an automated approach employing DMTCP and Slurm is adopted through the deployment of a single job script. This script consolidates both checkpointing and restarting functionalities, reducing the need for manual oversight. It adeptly monitors the maximum compute time, manages job requeuing, appends output/error files, and captures signals for timely checkpointing. The integration of Slurm \verb|--comment| and \verb|--signal| flags enables precise tracking of compute time and initiates checkpointing prior to the job reaching its time limit. This streamlined automation fosters a seamless operational cycle, ensuring jobs can progress from their latest checkpoint without necessitating a restart from the beginning, thus markedly elevating both efficiency and the reduction of computational redundancies~\cite{nersc-doc}.

\subsection{At NERSC Perlmutter inside the Containers}
Recognizing the widespread use and replicability benefits of the NERSC containers; 
{\tt shifter}\footnote{\href{https://github.com/NERSC/shifter}{https://github.com/NERSC/shifter}} 
and {\tt podman-hpc}\footnote{\href{https://github.com/NERSC/podman-hpc}{https://github.com/NERSC/podman-hpc}}, 
we opt to utilize them for testing C/R jobs with DMTCP. 
However, we face a limitation: DMTCP can not perform a checkpoint from outside the container; 
it has to be included within the container at the time of its creation. 
The construction of the simulation package within the containers proves to be versatile, 
as it could be accomplished during the container's build process or after the container 
was constructed by linking the source code from an external location. 
Additionally, it was possible to augment the container's capabilities by building upon 
an existing container, which facilitates rapid experimentation with minimal alterations. 
All these methods underwent thorough testing and validation. 
The following Dockerfile code demonstrates how to integrate DMTCP into an existing container with minimal modifications:
\\
\begin{verbatim}
FROM my_application_container:latest
RUN git clone 
https://github.com/dmtcp/dmtcp.git \
    && cd dmtcp \
    && ./configure && make \
    && make install
\end{verbatim}
This snippet illustrates the necessary commands to retrieve the latest DMTCP source from the repository and compile it as part of the container's setup, ensuring that it is embedded within the container environment from the outset. Furthermore, in the context of Geant4 simulations, we leverage the CernVM File System (CVMFS)~\cite{P_Buncic_2010}, which conveniently provides access to various Geant4 versions, thus simplifying the testing and deployment process across multiple versions.

To enhance operational efficiency in containerized HPC settings, we developed a system for both manually and automatically submitting and resubmitting jobs using DMTCP's C/R features. This approach uses scripts that set job properties and activate the C/R functions in {\tt shifter} and {\tt podman-hpc} containers at the NERSC Perlmutter cluster. Significant modifications have been implemented in the {\tt shifter} container script to ensure compatibility with {\tt podman-hpc} and vice versa.

\subsubsection{Automated C/R Strategies}

\begin{figure}[h]
	\centering
    \includegraphics[width=0.45\textwidth]{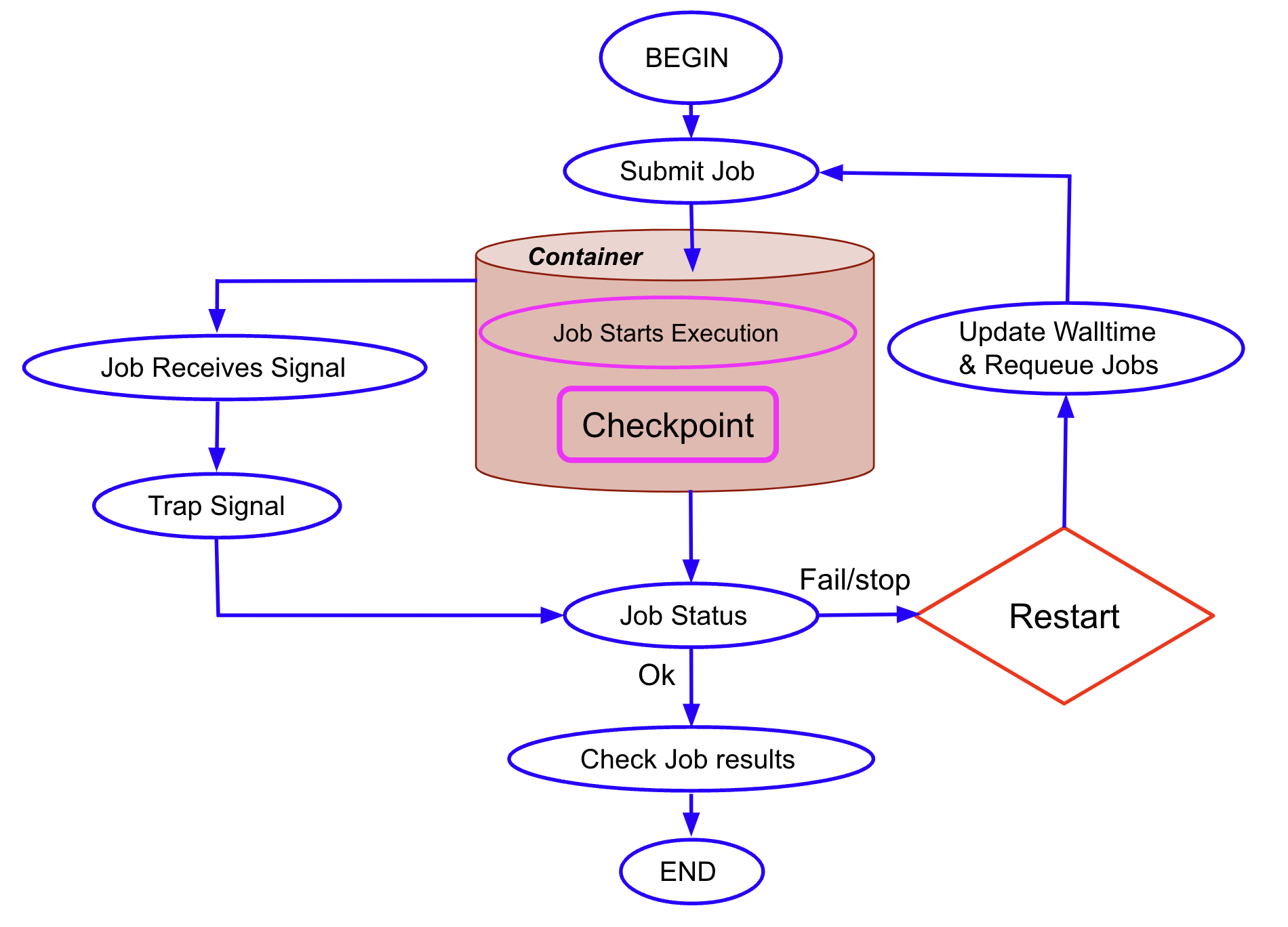}
     \caption[]{Operational workflow of automated job management in the NERSC containerized HPC environments. This figure delineates the end-to-end process flow within a containerized HPC environment, encapsulating job submission, execution, checkpointing, signal trapping, and the conditional logic for job restarting or completion. It serves as an illustrative guide to the DMTCP-enabled job resubmission mechanism.}
    \label{CR_flowchart}
\end{figure}

A workflow diagram, as presented in Fig.~\ref{CR_flowchart}, visually communicates the automated job management process, detailing the operational flow from job start to finish within a containerized HPC context. This diagram delineates each step from submission through execution, covering signal detection and the activation of the embedded C/R mechanism in NERSC containers. The decision paths post-receipt of a termination signal, showcase either completion checks or job failure handling leading to potential restarts.

A batch script handles the nuances of job management. It includee robust functions for administration and monitoring of SLURM~\cite{osti_15002962} jobs. A custom script, integrating time tracking, signal trapping, job requeuing, and DMTCP for C/R functionality, is designed and included in the batch script. This script is particularly pivotal in converting execution time into a human-readable format, calculating the remaining time for job scheduling, updating job comments to reflect the current status, and managing job requeuing based on the calculated remaining time. Furthermore, it integrates with DMTCP to facilitate C/R functionalities, ensuring that such operations are seamlessly woven into the fabric of job management and monitoring workflows.

Users initiate their computational tasks with batch scripts that include DMTCP within the container, ensuring compatibility with essential software packages like Geant4 and CP2K~\cite{Khne2020}. The above-mentioned custom script is carefully crafted to circumvent the constraints of container environments where direct intervention for C/R tasks is not inherently possible. It employs a \verb|restart_job| function that integrates a \verb|start_coordinator| to launch the checkpointing mechanism, followed by the execution command \verb|dmtcp_launch| for efficient job lifecycle management.

The robustness of the process is further fortified by handling termination signals such as \verb|SIGTERM|. The system is configured to intercept these signals, thereby triggering a \verb|requeue| function, which resubmits the job. This automation ensures a continuous execution flow and optimal resource utilization. Furthermore, specific Slurm directives are tailored for the automatic resubmission of C/R jobs with DMTCP. These directives, annotated within the user's job script, define the minimum time allocation, signal handling for termination, and rules for job requeueing, thus providing a fail-safe against premature terminations. The output file handling is configured to append, allowing for a seamless continuation of logging.

The technical implementation involves setting an environment variable \verb|DMTCP_COORD_HOST| to facilitate the restart of jobs. This is followed by a defined function \verb|requeue| within the script that echoes the job's status and instructs Slurm's control mechanism to requeue the job using the job ID. A signal trap for \verb|SIGTERM| is also set up to activate this requeue function.

Ultimately, the job is encapsulated within the NERSC container through a series of commands that ensure the DMTCP coordinator is operational, and the job can commence within the controlled environment. A trap for checkpointing on termination signals is established, and the \verb|restart_job| function is executed to manage the job's lifecycle. This layered approach culminates in a resilient and automated workflow. It guarantees not just the execution but also the seamless resumption of jobs in response to interruptions, thus aligning with the dynamic demands of modern HPC workloads.
\\
\subsubsection{Manual C/R Strategies}
In addition to the automated systems, the manual submission and resubmission of jobs hold significant importance, especially in scenarios where detailed monitoring and specific control over the C/R process are required. For the manual process, the job is first submitted using the same function as in the automated process, which creates a checkpointing state. This initial submission is crucial as it establishes a reference point from which the job can be restarted in case of failure or interruption. After the job commences, the user actively monitors its output and error logs to identify any anomalies or points of failure. Based on this analysis, the user can decide whether to resubmit or restart the job.

The manual intervention process involves utilizing a file created during the checkpointing phase. This file serves as a snapshot of the job’s state at a particular point in time, allowing for a precise restart from that state. The user manually intervenes to restart the job using this checkpoint file, effectively resuming the process from where it left off. This iterative process of submission, monitoring, checkpointing, and manual restart continues until the job successfully completes its execution. 

This manual approach provides an added layer of control, enabling users to fine-tune the job execution process and address specific issues that may not be automatically detectable. It complements the automated system, offering a comprehensive solution for managing jobs within the NERSC containerized HPC environment, ensuring both flexibility and reliability in handling complex computational tasks.

The complete code for all the above-mentioned methods can be found in~\cite{madantimalsina_checkpoint_restart_dmtcp_g4}.
%%%%%%%%%%%%%%%%%%%%%%%

%%%%%%%%%%%%%%%%%%%%%%%
\section{Results}
To evaluate the robustness and versatility of our checkpoint-restart (C/R) mechanism, we conducted a series of tests across a spectrum of Geant4 versions~\cite{AGOSTINELLI2003250}, namely 10.5, 10.7, and 11.0, while engaging various simulation environments. These environments were diverse, encompassing electromagnetic (EM) calorimeter arrays, hadron sandwich calorimeters, and specialized water phantom simulations designed for voxel geometries~\cite{git_km}. Additionally, we expanded our investigation to scenarios featuring both neutron and gamma-ray sources, thereby encompassing a broad range of particle interactions. Our examination delved into neutron measurement and characterization simulations, employing a variety of sources such as AmLi, AmBe, and Cf-252, and utilizing a Helium-3 proportional counter for detection purposes. Furthermore, we conducted simulation tests for the characteristic study of gamma emissions from various isotopes, including Na-22, K-40, and Co-60, employing High Purity Germanium (HPGe) detectors to capture the data~\cite{sharma2022calibrations}.

The breadth of these tests was designed to rigorously assess the checkpoint-restart process's ability to handle preemptions seamlessly. The positive outcomes of these tests—where each job, regardless of the simulation complexity or nature, was preempted, subsequently resumed, and brought to successful completion—provide compelling evidence of the resilience and reliability of our C/R process. This series of tests has underscored the efficacy of our implementation across a diverse array of simulated environments, reinforcing the practicality of C/R mechanisms in facilitating uninterrupted and resilient computational research.
%%%%%

\begin{figure}[!htb]
    \centering
    \includegraphics[width=\linewidth]{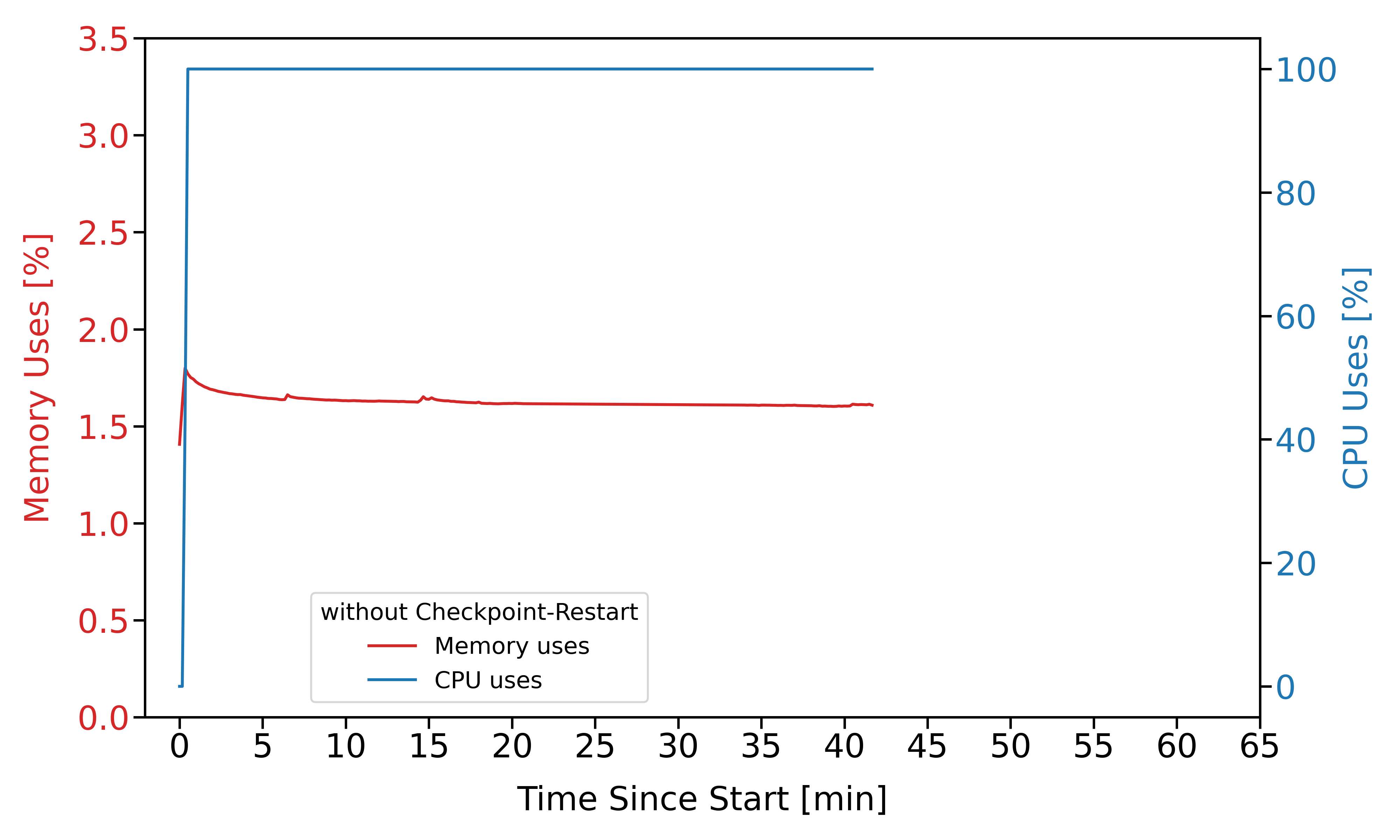}
    \includegraphics[width=\linewidth]{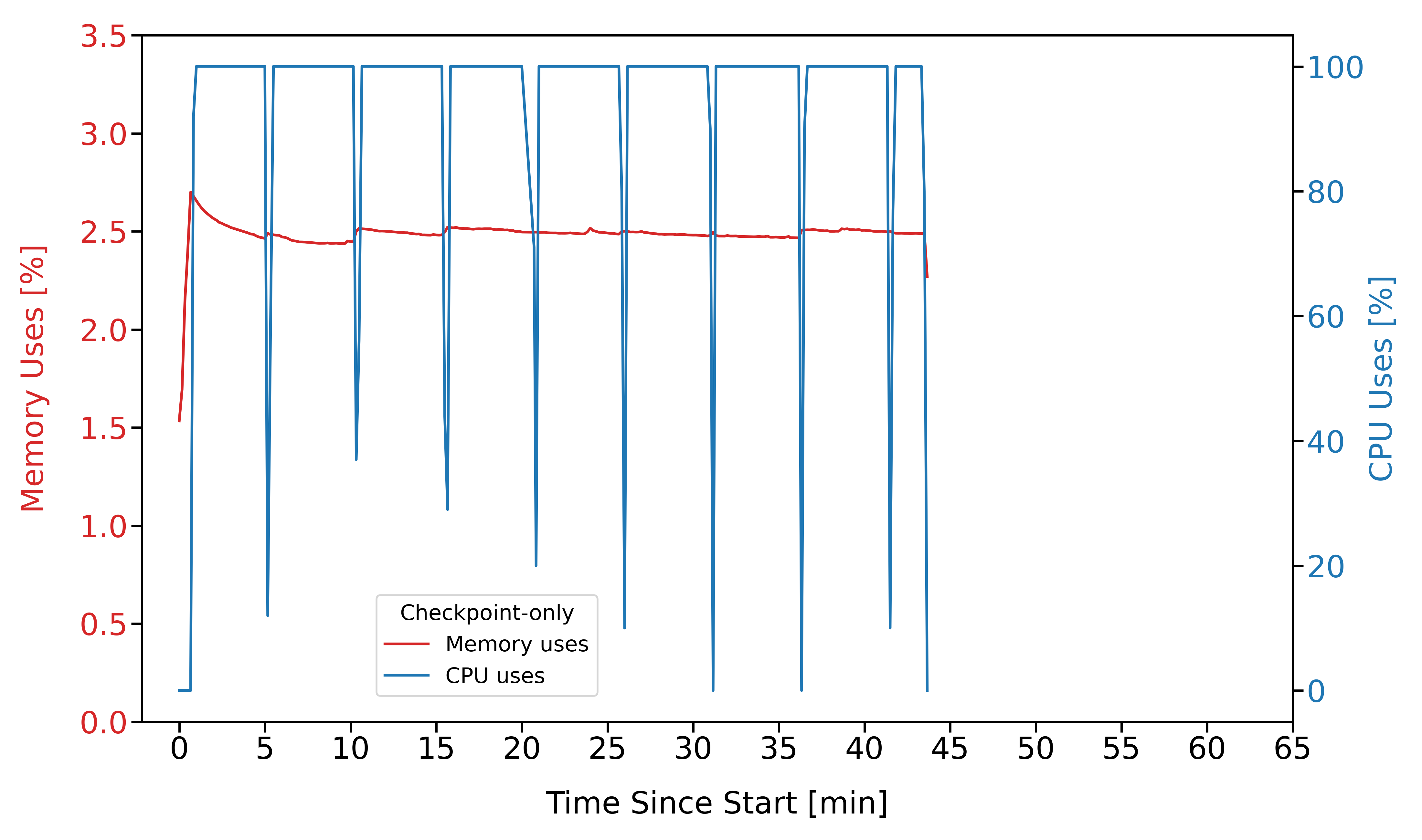}
    \includegraphics[width=\linewidth]{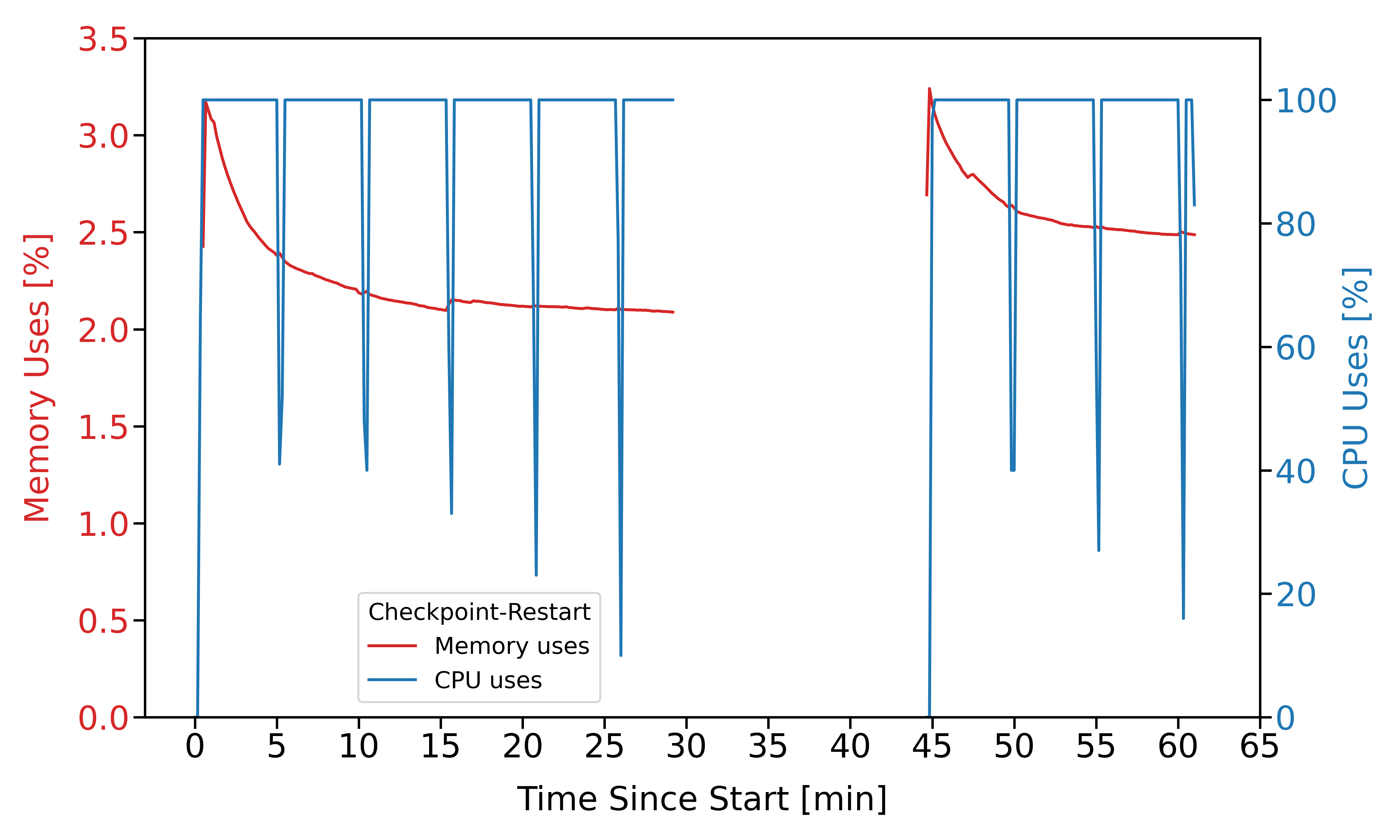}
     \caption[]{Comparative analysis of memory and CPU utilization over time at NERSC Perlmutter for computational processes using different strategies: without checkpoint-restart (top), checkpoint-only (middle), and with checkpoint-restart (bottom) within {\tt shifter} container.}
    \label{CR_result}
\end{figure}

In pursuit of evaluating the efficacy of our C/R implementation, we orchestrated comparative computational runs across three distinct scenarios: without checkpoint-restart (C/R), with checkpoint-only, and with both checkpoint and restart. These experiments were conducted to assess the consequent effects on the total runtime and memory footprint of the processes, employing NERSC’s {\tt shifter} and {\tt podman-hpc} containers. The resulting data, as depicted in Fig.~\ref{CR_result}, reveal the resource utilization patterns intrinsic to each strategy. 

Without the integration of C/R mechanisms, the upper plot showcases the system’s behavior under a normal operational regime. The memory usage, depicted by the red line, experiences a marked initial drop and then continues to decrease gradually (some spikes in-between may be due to other processes running on the system background), which could be indicative of the system's effective memory management or a diminishing workload. Concurrently, the CPU utilization, shown by the blue line, remains relatively constant, implying a consistent use of computational resources without significant fluctuations that might otherwise indicate process interruptions or variable computational demands.

Moving to the middle plot, which represents the \say{checkpoint-only} scenario, the data exhibits a distinctly different pattern. Here, we observe sharp increases in memory usage at regular checkpoint intervals. These peaks in memory usage reflect the additional memory required to capture the system's state, a characteristic burden of the \say{checkpointing} process. Corresponding with these spikes in memory, there are deep declines in CPU utilization. These drops suggest that the system momentarily diverts computational efforts from its primary tasks to handle the \say{checkpointing} process, effectively pausing other activities.

In the bottom plot, illustrating the \say{checkpoint-restart} condition, we observe significant memory usage spikes at each checkpoint, indicative of the additional memory allocation required for checkpointing action. Subsequently, memory usage diminishes steadily, suggesting efficient memory deallocation. The CPU usage demonstrates a decline following each checkpoint, with intermittent peaks corresponding to the computational load during the checkpointing process. This gap between the 29th and 45th minutes as expected, is associated with job preemption and batch job requeuing period as they await resource re-allocation for restart. 
%The concurrent decrease in CPU usage and increase in memory suggest a state of computational inactivity, where the process is in queue for reassignment.
Beyond the 45th minutes, the process resumes on a new node, and checkpointing activity recommenced.

The graphical analysis provides insight into the operational efficacy of varying C/R strategies. In the absence of C/R measures, depicted in the top graph, the task is completed quickly and with lower memory use, setting a benchmark for optimal performance. The middle graph indicates that incorporating checkpoints moderately extends task duration (by a few minutes) and increases memory demands ($\sim 0.8 \% $) due to the overhead of state preservation. The bottom graph reveals the preemption of jobs and their subsequent resubmission by the batch scheduler, leading to longer job completion times.
%akin to the middle scenario. 

Comparative computational analyses indicate that C/R techniques exhibit a slight increase in memory usage due to the loading of DMTCP and associated files. However, the pivotal advantage of the 'checkpoint-restart' strategy, as represented in the bottom graph, lies in its capability to autonomously requeue and resume from the last saved state. This feature is vital, as it circumvents the need to restart jobs from the initial state, substantially diminishing both the temporal and resource expenditures typically incurred. This strategic efficiency, which we have repeatedly highlighted in our paper, showcases the significant cost and time savings enabled by our C/R implementation, as validated by the depicted resource utilization trends.

The data for the aforementioned plots were acquired using the Lightweight Distributed Metric Service (LDMS) provided by Sandia National Laboratories, and were processed utilizing code from the OVIS-HPC~\cite{7013000_Age,ovis-hpc}. This suite of plots collectively provides an empirical basis for understanding the trade-offs and operational dynamics associated with different checkpointing techniques. It becomes evident that while checkpointing introduces additional overhead in terms of memory and CPU cycles, it is an indispensable mechanism for ensuring data integrity and system resilience, particularly in long-running computations. The distinct patterns observed affirm the stability of computational load when uninterrupted by checkpointing and highlight the impact of our C/R implementation on preserving computational continuity at NERSC.

\section{Future Directions}
Our focus has been on Geant4-based simulations, particularly beneficial for HEP and medical science. We are expanding our research to include other domains like material science. Tests with CP2K are ongoing; so far, we've made progress with checkpointing, although we have encountered some issues with restarting. We are collaborating with the developers of DMTCP and CP2K to address these problems. In the near future, we intend to expand our testing to include a broader range of material science software, such as VASP, BerkeleyGW, LAMMPS, and others.

In addition, we aim to explore the utilization of MANA (MPI-Agnostic Network-Agnostic)~\cite{garg2019mana, xu2023implementationoblivious} for our checkpointing needs, building upon our existing foundation with DMTCP. MANA promises enhanced efficiency and flexibility for MPI applications through its innovative split-process approach, which simplifies the checkpointing process by focusing on application state while abstracting away MPI library and network specifics. This strategic shift is expected to significantly bolster our capability in managing complex computational workflows in a more MPI-centric environment.
%%%%%%%%%%%%%%%%%%%%%%%

\section{Related Work}
\label{sec:related_work}
In the realm of HPC, a range of C/R technologies and container solutions have been developed, each with its unique focus and applications. This section compares these technologies with our use of DMTCP in {\tt shifter} and {\tt podman-hpc} containers at NERSC, highlighting how our approach fits within the broader landscape of HPC innovations.

We recognize similar studies with related themes, but they differ in their approach and focus. There are studies on container technologies that focus on live migration in RDMA applications~\cite{273765}, and others examining various container technologies and their performance~\cite{10.1007/978-3-030-63089-8_9}, which differ from our approach in HPC environments using DMTCP. Our research is distinct in its practical implementation and testing of DMTCP in specific computational tasks and simulations, rather than an overarching evaluative methodology. In the studies, Docker in fog computing with DMTCP~\cite{9126743}  the primary difference lies in the application domain (fog computing vs. HPC) and the specific problem being addressed (container deployment efficiency vs. computational efficiency in HPC environments). Although both BLCR's system-level checkpointing on Linux clusters~\cite{Paul_H_Hargrove_2006} and our method address checkpoint-restart processes in HPC, there is a key difference between them. BLCR uses system-level implementation, whereas DMTCP uses application-level implementation. Additionally, the specific technologies and methodologies employed in both methods, while relevant, diverge from our focus. Furthermore, security requirements are not compatible with Singularity's containerized solutions in HPC. Therefore, we place our unique emphasis on optimizing checkpoint-restart processes in HPC with DMTCP.

In summary, our study uniquely positions DMTCP within the NERSC's {\tt shifter} and {\tt podman-hpc} framework, considering the specific research requirements and security protocols. This differentiates our work from other C/R solutions and container technologies, underlining its innovation and significance in advancing HPC computational methodologies.

%%%
%%%%%%%%%%%%%%%%%%%%%%%%%%%%%%%%%%%%%%%%%%%%%%%%
\section{\bf Conclusion}
\label{sec:CONCLUSIONS}
This study effectively demonstrates the utility of checkpoint-restart techniques using DMTCP in HPC, both inside and outside containers. Our approach is especially valuable in HPC environments where computations are complex and lengthy, as it significantly reduces the cost and time associated with restarting processes from scratch. By successfully implementing this strategy in various simulations, including those crucial to high-energy physics, medical science, and ongoing research in material science, this research underscores a critical advancement in efficient and reliable computational methodologies. Our findings not only affirm the technique’s effectiveness in real-world applications but also open avenues for broader applications in computational science, highlighting the potential for further innovation and optimization in HPC processes.
%%%
%%%%%%%%%%%%%%%%%%%%%%%%%%%%%%%%%%%%%%%%%%%%%%%%
%%%%%%%

%%%
%%%%%%%%%%%%%%%%%%%%%%%%%%%%%%%%%%%%%%%%%%%%%%%%
%%%%%%%
\section*{ Acknowledgments}
\label{sec:Acknowledgments}
This research used resources of the National Energy Research Scientific Computing Center (NERSC), a Department of Energy Office of Science User Facility.
The authors also wish to acknowledge the technical input and support provided by the NERSC DAS and DESG groups. Special thanks to Wahid Bhimji, the lead of the DAS group, as well as Dhruva Kulkarni from the APG group for their invaluable contributions.

\footnotesize
\bibliographystyle{unsrt}
\bibliography{bibliography_file} 
%\bibliographystyle{IEEEtran} % or another style you prefer
%\bibliography{bibliography_file} % the name of your .bib file without the .bib extension
\normalsize

\end{document}